\documentclass[10pt,conference]{IEEEtran}
\usepackage{algorithmicx, algorithm}
\usepackage{algpseudocode}
\usepackage{float}
\usepackage{caption}
\usepackage{subcaption}
\usepackage{xcolor}
\usepackage{mathtools}
\usepackage{fixltx2e}
\usepackage{array}
\usepackage{amssymb}
\usepackage{multirow}
\usepackage{hhline}
\usepackage{balance}
\usepackage[backend=bibtex,style=numeric,citestyle=numeric,maxbibnames=99]{biblatex}
 
\usepackage{paralist}

\newcommand\acro{{\sf FADEWICH}}
\newcommand{\lname}{{Fast Deauthentication over the Wireless Channel}}

\newcolumntype{?}{!{\vrule width 1pt}}
\newcommand{\specialcell}[2][c]{%
\begin{tabular}[#1]{@{}c@{}}#2\end{tabular}}

\begin{document}
 
\title{FADEWICH: Fast Deauthentication\\over the Wireless Channel}

\author{
\IEEEauthorblockN{
	Mauro Conti\IEEEauthorrefmark{1},
	Giulio Lovisotto\IEEEauthorrefmark{1}\IEEEauthorrefmark{2},
	Ivan Martinovic\IEEEauthorrefmark{2},
	Gene Tsudik\IEEEauthorrefmark{3}
}
\IEEEauthorblockA{
\IEEEauthorrefmark{1}University of Padua, IT
    \\conti@math.unipd.it, giulio.lovisotto@studenti.unipd.it}
\IEEEauthorblockA{\IEEEauthorrefmark{2}University of Oxford, UK
    \\\{giulio.lovisotto, ivan.martinovic\}@cs.ox.ac.uk}
\IEEEauthorblockA{\IEEEauthorrefmark{3}University of California, Irvine, US
    \\gene.tsudik@uci.edu}
}

\maketitle

\begin{abstract}
Both authentication and deauthentication are instrumental for preventing unauthorized 
access to computer and data assets. While there are obvious motivating factors for
using strong authentication mechanisms, convincing users to deauthenticate is not straight-forward,
since deauthentication is not considered mandatory. A user who leaves a logged-in workstation 
unattended (especially for a short time) is typically not inconvenienced in any way; in fact, the 
other way around -- no annoying reauthentication is needed upon return. However, an unattended 
workstation is trivially susceptible to the well-known ``lunchtime attack'' by any nearby adversary 
who simply takes over the departed user's log-in session. At the same time, since deathentication
does not intrinsically require user secrets, it can, in principle, be made unobtrusive.
To this end, this paper designs the first automatic user deauthentication system -- \acro\ -- that does not rely on
biometric- or behavior-based techniques (e.g., keystroke dynamics) and does not require users to 
carry any devices. It uses physical properties of wireless signals and the effect of human bodies on their 
propagation. 

To assess \acro's feasibility and performance, extensive experiments were conducted with 
its prototype. Results show that it suffices to have nine inexpensive 
wireless sensors deployed in a shared office setting to correctly deauthenticate {\bf all}
users within six seconds (90\% within four seconds) after they leave their workstation's vicinity.
We considered two realistic scenarios where the adversary attempts to subvert \acro\
and showed that lunchtime attacks fail.
\end{abstract}

\section{Introduction}
To prevent unauthorized access to various restricted resources, most computer systems mandate  
authentication and deauthentication mechanisms. Unfortunately, their efficacy is weakened 
since many users are too lazy, too distracted, or simply annoyed by these security procedures.
Users who do not care about protecting resources (because they do not understand either
their vaue or seriousness of threats) often attempt to avoid, circumvent or simplify
security procedures, e.g., select easy-to-remember  passwords~\cite{ur2015added}
that are also easy to crack~\cite{bonneau2012science}. 

In most multi-user settings (e.g., home, office, school) most users tend to, perhaps grudgingly, 
accept the need for authentication. Moreover, mandatory system rules can dictate  
selection and change criteria, such that trivial passwords are avoided and new passwords
are periodically required. However, since deauthentication requires no passwords (or any other
secrets) easily annoyed, lazy or absent-minded users can leave their log-in session unattended,
for numerous reasons, e.g., make a private phone-call, have a coffee, use the restroom
or go to lunch. In this case, any physically nearby adversary can easily perform a so-called 
``lunchtime attack''~\cite{Eberz2015}, which basically means: walk up to the unattended computer
and gain access to the current log-in session of the authorized departed user.

A great deal of prior research has been devoted to user authentication, yielding many
techniques  some of which are both effective (i.e., unavoidable), and usable, even 
transparent~\cite{al2015survey}. Currently, the most popular form of authentication involves
providing traditional account credentials, i.e., username and password. At the same time, 
biometric-based techniques (e.g., fingerprint, voice, writing, iris, and face recognition) are 
gaining popularity both as stand-alone or second-factor means of authentication.

In contrast, much less attention has been paid to deauthentication.  In particular, there is no 
effective (i.e., with low error rates) user-transparent deauthentication method.
To the best of our knowledge, other than mandatory re-login after a fixed interval of user-input inactivity,
there is no widely used deauthentication method. Unfortunately, if this interval is too short, users are
annoyed by having to re-authenticate often and perhaps unnecessarily. Meanwhile, 
if it is too long, lunchtime attacks become more likely. In fact, lunchtime attacks are always possible
when inactivity intervals are used: interval size determines the adversary's {\em window of opportunity}.

Alternatively, explicit deauthentication can be mandated. It can be made very easy, e.g., just 
a single mouse click on a screen-lock icon. However, lazy, careless or distracted users
neglect to follow the rules~\cite{Sinclair2012, Sinclair2008}.
Since username/password-based authentication will remain in wide use for the foreseeable
future~\cite{Herley2012, Shay2014}, automatic deauthentication remains an open challenge.
An ideal deathentication method would be unobtrusive (user-transparent) and highly effective.

Although lunchtime attacks involve physical constraints (i.e., the adversary must be physically  
near the victim), this is not an excuse to ignore the problem. In fact, insider 
attacks are very common and potentially very dangerous. For example, a 2014 survey of 557 
large organizations showed that over half (53\%) experienced an insider-caused security
incident~\cite{CERTInsiderThreatCenter2014}. Also, average financial loss caused by such an 
incident exceeds US\$100,000 ~\cite{insiderthreatstudy}. Furthermore, as reported in 2015 by 
Verizon~\cite{verizon}, \textit{insider and privilege misuse} are the most frequent types of
security incidents among 9 incident categories. Typically, malicious insiders exploit multiple opportunities 
throughout  working hours to attempt to gain unauthorized access. Plus, being usually aware of 
local system vulnerabilities, they can act very quickly.

In this paper, we design a new deauthentication method, called \lname\ (\acro), which 
uses multiple wireless (WiFi) sensors 
deployed within an office. These devices communicate through the wireless radio channel and 
monitor physical properties of this channel. For actual deauthentication we take advantage of 
the effect that human bodies have on the propagation of high frequency wireless signals. Specifically, 
when a person crosses (or stands in) the path between a transmitter and a receiver, the body affects 
wireless signal propagation, such that the receiver measures the transmitted signal with a different strength, 
i.e., Received Signal Strength Indicator (RSSI) changes~\cite{Zhao2013, Wilson2010}.
A receiver determines signal strength of a wireless signal emitted by a transmitter by 
combining received signal components. In cluttered indoor environments 
(e.g., a typical office setting), signal strength is determined by components that arrive at the receiver 
after being scattered, reflected and diffracted by obstacles, such as walls and objects 
(\textit{multipath propagation}). A person moving near wireless sensors causes signal components to 
change, producing fluctuations in measured signal strength, primarily by breaking the line-of-sight (LoS)
condition between them, and by altering propagation of multipath signal components~\cite{Wilson2012}. 

\subsection*{Contributions}
This paper's main contribution is a new deauthentication method -- \acro\ -- suitable for a 
typical multi-user or shared
office setting, where computers and workstations are not physically protected, e.g., by doors.
In such a setting, a malicious insider can attempt to gain physical access to a workstation (or computer)
that is left logged-in and unattended. \acro\ is easy to deploy, user-transparent and does not
require users to have any extra devices.
\acro is also efficient -- it deauthenticates a user within a few seconds after stepping away -- 
and usable, i.e., even idle users who remain at the workstation are not deauthenticated.

To assess feasibility and efficacy of \acro, we implemented it and evaluated its 
performance in a realistic office setting. We designed and ran experiments with fewest possible
assumptions: while users were aware of the system, they did not interact with it directly.
Also, they were asked not to change their normal behavior, meaning that there are no 
limitations on how and how frequently they moved. We identified two realistic insider 
lunchtime attack scenarios  We show that, with enough sensors, \acro\ prevents all such
lunchtime attacks. Moreover, we compare \acro\ with simple time-out-based 
deauthentication and show that loss of usability in the former is negligible, especially 
considering security benefits that it offers.

\section{Related Work}\label{sec:rel_works}
We now overview wireless signal strength analysis for localization purposes 
(Section~\ref{ch:rssi_processing}), followed by state-of-the-art authentication and 
deauthentication techniques (Section~\ref{ch:auth_deauth}).

\subsection{RSSI for localization}\label{ch:rssi_processing}
Researchers have been using the effect of the human body on wireless signals for several years. Bahl et al. in their pioneering RADAR~\cite{Bahl2000}, used RSSI processing for localization purposes. RADAR was among the first works to consider the obstruction effect of the user body on the signal strengths, and thus point out the difference of the measured values between transceivers in LoS and not-LoS condition. 

In wireless sensor networks, RSSI processing has been used for device free localization, where people can be located inside the monitored environment even if they do no carry a wireless enabled devices, using the body obstruction effect on the signals~\cite{kosba2012rasid}. 
Kaltiokallio and Bocca~\cite{Kaltiokallio2011} used this approach for intrusion detection.
The system in~\cite{Kaltiokallio2011}, is able to detect and track the presence of an intruder who is moving inside the area monitored by wireless sensors. 
Further improvements were made, and RSSI processing was combined with Radio Tomographic Imaging (RTI) in order to track  multiple people~\cite{Bocca2013,Zhao2013}. 
RTI use a large number of wireless sensors to tackle the unpredictable nature of the radio environment: the information from several pairs of sensors is combined in order to infer the presence and movement of the person. 

Although RTI techniques proved to be very effective to track people locations, these are not applicable in our context. 
In fact, in very cluttered and small rooms, the multipath components resulting from the reflections due to the walls, the objects, and the physical presence of users, have a fundamental role in the observed signal strength, producing very noisy and unpredictable changes~\cite{Wilson2012}.
Moreover, RTI is based on an initial \textit{calibration} phase, where the system understands the behavior of the radio environment when there is nobody in it. 
Afterwards, the fluctuations caused by the movement of users result different compared to the learned behavior. 
In our model a static and long-term calibration is not possible, since there is not an unique ``steady state'': the environment is dynamic, users may be walking inside the office, standing still, or sitting in their chairs
.

\subsection{Deauthentication Solutions}\label{ch:auth_deauth}
Since the deauthentication strongly depends on the authentication scheme, an overview on these authentication mechanisms is necessary.
Nowadays, the most popular authentication scheme is the use of passwords. 
Using passwords for the authentication has several advantages: they are very intuitive and convenient to use, users do not need to perform any activity (after the log in) nor carry any device. 
However, one of their main drawbacks is that there is no way of automatically deauthenticate users, exposing the user accounts and the system to possible threats.
Security researchers have thoroughly pointed out the weaknesses of password-based authentication~\cite{bonneau2012science, Shay2014,BonneauHerley2012,Herley2012}, and they have invested lots of effort in the search for usable and secure alternatives.

In this direction, a promising field is behavioral biometrics. 
Behavioral biometrics are typically passive (or \textit{transparent}) to the users, and can be used for continuous authentication.
With continuous authentication techniques, the user is continuously authenticated as long as he interacts with the system, and he is deauthenticated once his interactions stop, providing automatic deauthentication.
One of the more popular biometrics in this category is keystroke dynamics~\cite{banerjee2012biometric}. 
Even if these biometrics reached low error rates~\cite{frank2013touchalytics}, and are approaching a wider adoption, their security guarantees remain questionable.
In fact, error rates does not thoroughly represent the security properties of a biometric system: they only shows the resilience against the \textit{zero-effort} attack (i.e., the success rate of one malicious user enrolled into the system that claims another user identity).
When the threat model changes, the error rates could become irrelevant.
For example, Meng et al.~\cite{Tey2013} have examined the reliability of the keystroke dynamics, and they showed that an adversary is able to reproduce one user behavior after a brief training (after observing the user typing on a keyboard).
Recently, more sophisticated biometrics like the ones based on the pulse response~\cite{Rasmussen2014}, or on the eye movements~\cite{Eberz2015, sluganovic2016using} have been investigated. 
Although these solutions are harder to observe and to craft by an adversary, their deployment requires specific hardware, which may be considerably expensive, and that hinders their large-scale adoption.

Another approach to the deauthentication, is to adopt techniques based on proximity.
In these techniques, users carry small tokens that communicate over a short range radio channel with the workstations, providing continuous authentication~\cite{corner2002zero,Stajano2011}. 
The workstations periodically poll the tokens via a secure channel to detect their proximity, and they are automatically secured as soon as the token is unreachable.
The major drawback of proximity based authentication is that it requires the user to carry some device (and the system administrator to deploy and manage them), exposing to serious threats in case such devices are stolen.

Mare et al. proposed ZEBRA~\cite{mare2014zebra}, an hybrid approach for the continuous authentication. 
In their work, they have users wear a bracelet with an in-built gyroscope on their dominant wrist, and they combine the inputs received on the workstation (i.e., mouse movement and keyboard typing) with the actions observed by the bracelet. 
Where the two time series of events do not correlate, the user is logged out and the workstation is locked.
However, as pointed out by Huhta et al~\cite{huhta2015pitfalls}, ZEBRA design is flawed: they showed that 40\% of attackers are able to remain logged in for more than 10 minutes by opportunistically choosing the time and type of interactions (bracelet and mouse movement, and keyboard typing).
The authors acknowledge that the system relies on assumptions that makes it vulnerable to attackers, and even if ZEBRA performs well against accidental misuse by innocent users, opportunistic attackers remain a tough challenge.

\section{System Model}\label{sec:system_model}
Our main goal is to automatically deauthenticate users who are logged into a 
workstation when they leave its proximity. The method must be inexpensive and 
easy to deploy with minimal (preferably none) hardware requirements. 
Another goal is usability which translates into user transparency and low error rates.
An error refers to a false positive, i.e., a user being mistakenly deauthenticated while still 
using, and being physically present at, the workstation, thus forcing a superfluous log-in (
as discussed in~\cite{Shay2014}, users find repetitive password entry time-consuming 
and very annoying). Also, the system must not invade user privacy, e.g., using a camera is 
obtrusive and generates privacy issues that might discomfit users~\cite{nouwt2005reasonable}).

The system model is as follows:
\begin{compactenum}
\item The environment is a workplace with $k$ workstations $w_1,$ $...,$ $w_k$. 
Users log in using some authentication mechanism the particulars of which are
not important for deauthentication. Each workstation monitors user input
activity and communicates the idle time to a central station. We assume that there is a single 
entrance -- the only way users can access and exit the office. 
\item There are $m$ wireless devices $d_1,$ $...,$ $d_m$ each 
capable of sending packets and monitoring signal strength of packets 
received from the others. This lets us obtain $m \times (m-1)$ streams of 
signal strengths, which are transmitted through a secure channel 
(not necessarily a wireless one) to a central station. 
\item At installation time, there is an initial  phase when the system automatically 
collects and labels data. No adversarial presence is assumed during this phase. 
\end{compactenum}

\subsection{Threat Model}
In the context of this paper, a threat means {\em unauthorized access to an honest user's account}. 
We assume that the adversary is potentially anyone who has physical access to the office, e.g., 
co-workers, supervisors, customers, janitors, and delivery personnel.  
The adversary's goal is to gain access to an active log-in session on the \textit{target} workstation
used by one \textit{victim} user. The adversary does not know that user's login credentials.
We distinguish between two sub-types of adversaries:
\begin{compactitem}
\item \textbf{Insider}: anyone with access to the outside of the office. This can be any employee of 
the same organization. However, insider is not supposed to enter the actual office housing 
the target workstation.
\item \textbf{Co-worker}: anyone with access to the inside of the office (e.g., a co-worker,
assigned to another workstation in the same office) who is not supposed to use the target workstation.
\end{compactitem}
Since workstations might be close to each other, co-worker can access the target faster than insider.
If automatic deauthentication is based on a fixed time-out, co-worker has a significant advantage. 

As for other types of attacks, Section~\ref{sec:resilience} discusses our reasons for not considering 
attacks that alter the physical propagation of wireless signals. We also do not consider {\em social 
engineering} attacks.

\section{System Design}\label{sec:system_design}
We now present technical details of \acro\ and motivate its design.

\subsection{Overview}\label{sec:overview}
The complete system includes several components and three main modules: Keyboard/Mouse Activity 
module (KMA, described in Section~\ref{sec:kma}), Movement Detection module (MD, 
described in Section~\ref{sec:md}), and Radio Environment module (RE, described in Section~\ref{sec:re}). 
These modules can access information provided by sensors and workstations, 
and a classifier that has been previously trained (see Section~\ref{sec:training_phase}).
Another component implements the control part of the system, which merges 
information provided by the modules to apply actions to the workstations.
Section~\ref{sec:overlap} discusses the overlapping movements problem, while 
Section~\ref{sec:rules} illustrates system actions and the workflow of \acro.

\begin{figure}[htpb]
    \centering
    \includegraphics[width=0.45\textwidth]{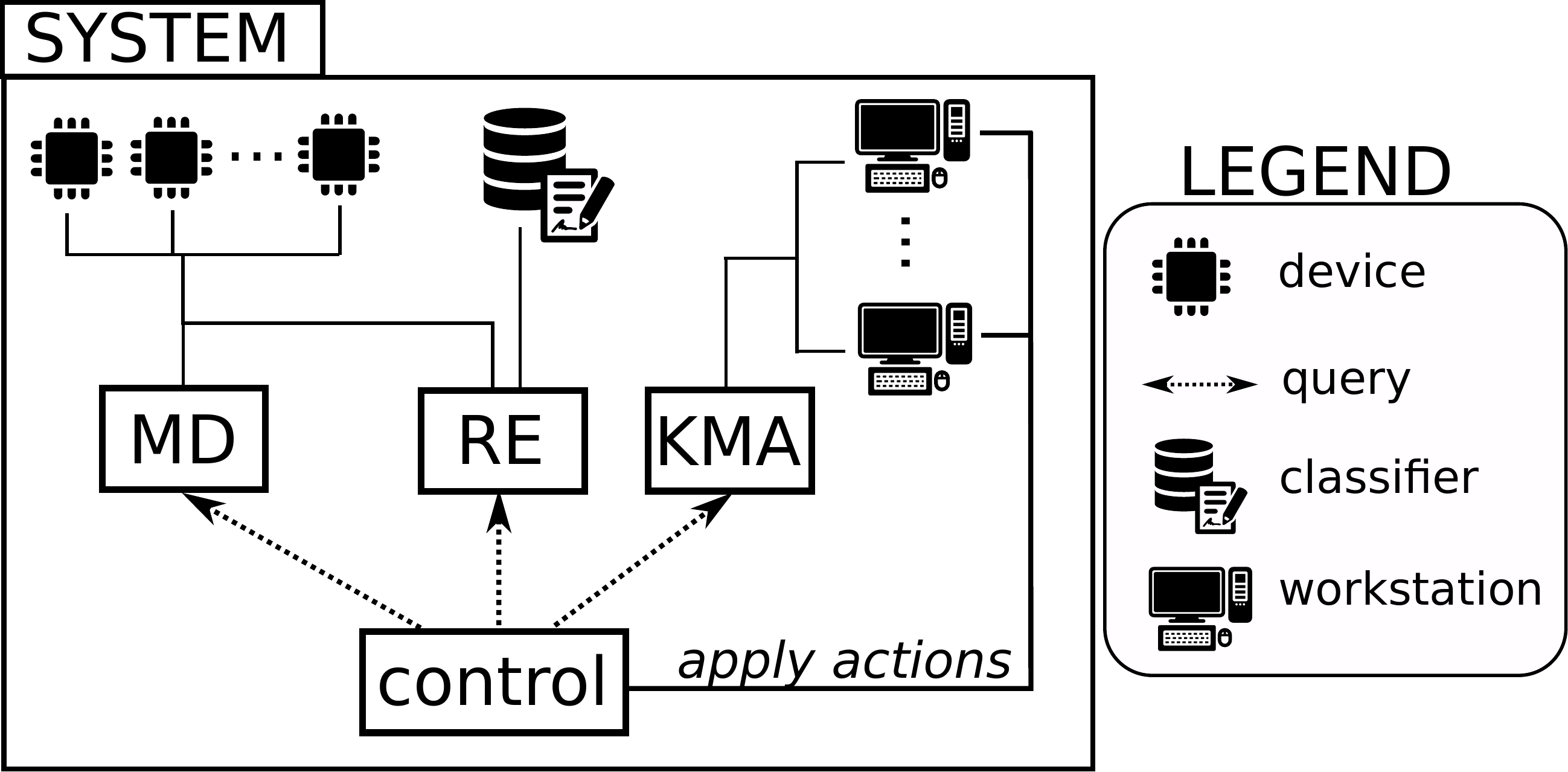}
    \caption{Overview of \acro\ components.}\label{fig:system_overview}
\end{figure}

\subsection{Keyboard/Mouse Activity (KMA)}\label{sec:kma}
This module monitors user input at each workstation, and keeps track of idle time, i.e., the 
interval of time the workstation observed no keyboard or mouse input. 
When the system at time $t$ asks KMA which workstations have been idle for $s$ seconds, KMA 
returns a set of workstations $S_{t}^{(s)}=\{w_h,\:...,\:w_j\}$ that have been idle between $t-s$ and $t$.

\subsection{Movement Detection (MD)}\label{sec:md}
This module obtains signal strength streams and determines whether there has been any  significant 
fluctuations in the radio environment. In the following, after introducing the notation,
we show how such fluctuations are identified based on the discrepancy with a learned 
level of fluctuations. 

\subsubsection{Notation}
We refer to signal strength measurement on stream $i$ at time $t$ as: $r_{t}^{(i)}$.
A window on a stream is a sequence of measurements, such that, for a given size $d$, and time $t$: 
\[ V_{t-d, t}^{(i)} = \{r_{t-d}^{(i)}, r_{t-d+1}^{(i)}, ..., r_{t}^{(i)}\}.\] 
Given a distribution $r = \{r_1, r_2, ..., r_n\}$, and a kernel function $K$, we refer to its estimated 
density function as:
\[\hat{f}_K(r) = \frac{1}{n h}\sum\limits_{i=1}^{n}K(\frac{r - r_{i}}{h}),\]
where $h$ is the bandwidth of the kernel.

\subsubsection{Standard Deviation Profile}
To recognize users' movements, we decided to use the sum of the standard deviations of signal streams. 
Initially, MD builds a distribution of these summations, using a sliding window to compute standard 
deviations of some period of data, e.g., 30 seconds in our experiments. At each time step $t$, the 
summation of standard deviations $s_t$ is computed as:
\[s_t = \sum\limits_{i}^{} \sigma_{V_{t-d, t}^{(i)}} \:\:,\]
where $\sigma_{V_{t-d, t}^{(i)}}$ is the standard deviation of measurements in window 
$V_{t-d, t}^{(i)}$, $d$ is window size, and $i$ identifies the stream. These values form a frequency distribution 
$s = \{s_0, s_1, ..., s_n\}$, that we refer to as the \textit{normal} profile. Density of distribution $s$ is 
estimated with a Gaussian kernel and the estimated function $\hat{s}$ is later used for comparison.

\begin{figure}[h!]
    \centering
    \includegraphics[width=0.45\textwidth]{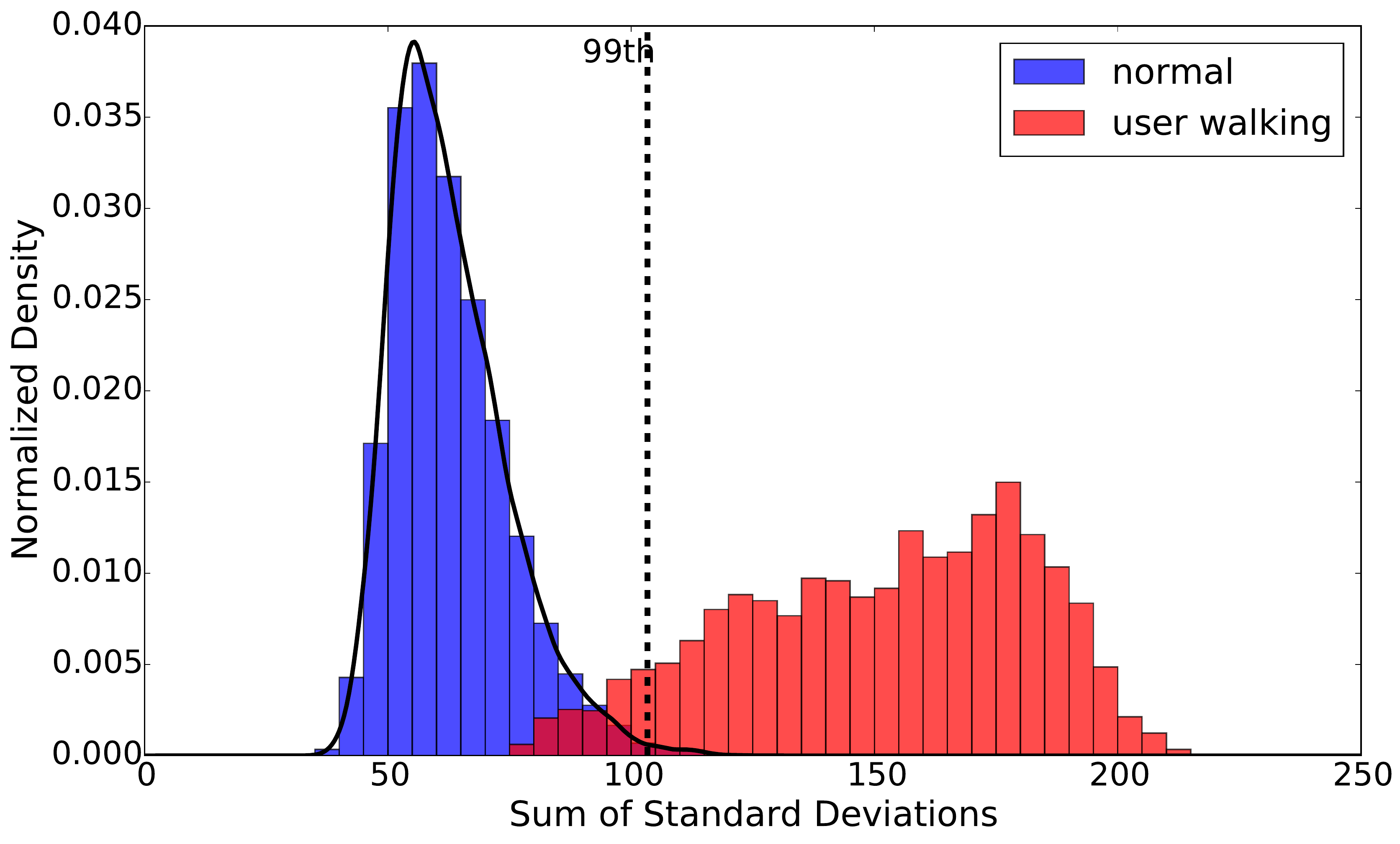}
    \caption{Frequency distribution of observed total standard deviation.}\label{fig:md_distrib}
\end{figure}

Thereafter, MD periodically computes the current sum of standard deviations $s_{t}$ with the latest observed
data. When $s_{t}$ exceeds the $(100-\alpha)^{th}$ percentile of cumulative distribution function $\hat{S}$
of estimated distribution $\hat{s}$, the module reports the anomalous changes observed.
When changes belong to the remaining part of the distribution, MD reports the normal state.
Figure~\ref{fig:md_distrib} shows the difference between the measurements of $s_t$ when no one is 
walking in the office, and when a user is walking inside. The solid line shows the probability 
distribution function estimated with the Gaussian kernel.

\subsubsection{Profile Update}
Due to the noisy nature of the radio environment, behavior of the streams varies slightly 
depending on several factors, in particular, the number of users in the room. 
To account for this phenomena, the normal profile needs to be updated with most recent measurements. 
We use batches of size $b$ for the update: each time current $s_t$ is computed, it is queued for the update. 
When the queue reaches size $b$, if less than a fraction $\tau$ of values in the queue belonged to 
$(100-\alpha)^{th}$ percentile of $\hat{S}$, they are added to the normal profile distribution 
by removing the oldest $b$ values. The kernel density estimation is performed again after each update.

Algorithm~\ref{alg:md_alg} shows MD workflow. Whenever MD returns \textit{anomalous}, 
the current state of the environment significantly differs from the learned profile. 
Whereas, if it returns \textit{normal},  the current state matches the profile.

\begin{algorithm}[t]
\small
  \caption{MD workflow}
  \label{alg:md_alg}
	\begin{algorithmic}[1]
	
	\State{ $\hat{s} \gets$  \textit{initialize\_normal\_profile()}}
	\State{ $Q \gets \{\}$ // batch for profile update}
	\While{True}
	\State $s_t \gets \sum\limits_{i}^{} \sigma_{V_{t-d, t}^{(i)}}$
	\State $ub \gets \hat{S}^{(100-\alpha)^{th}}$
	\State {$Q \gets Q + \{s_t\}$}
	\If {$s_t \geq ub$}
		\State {\textbf{return} \textit{anomalous}}
	\Else
		\If {$|Q| \geq b$}
			\If {\textbf{not} \textit{is\_anomalous(Q, $\tau$)}}
				\State{$\hat{s} \gets update\_distribution(\hat{s}, Q)$}
			\Else
				\State{$Q \gets \{\}$}
			\EndIf
		\EndIf
		\State {\textbf{return} \textit{normal}}
	\EndIf
	\EndWhile
	\end{algorithmic}
\end{algorithm}

\subsubsection{Variation Windows}
Hereafter, we refer to \textit{variation windows} as time intervals $[t_1, t_2]$, where MD has recognized
anomalous fluctuations that started at time $t_1$ and  continued until $t_2$, when the the environment 
went back to normal. Unfortunately, user movements are not the only cause of fluctuations. The 
radio environment is subject to other uncontrolled changes that may result in variation windows 
even if no one is moving.  To account for this possibility and to exclude other brief variations 
due to users moving slightly while remaining at their workstations, there is a threshold on the duration 
of variations windows $t_{\Delta}$. This way, variation windows shorter that $t_{\Delta}$ are ignored, 
while longer windows are interpreted as user movements and trigger a system decision. 
The value of $t_{\Delta}$ must be chosen very carefully, since it bears much impact on
overall performance; see Section~\ref{sec:md_performance}.

\subsection{Radio Environment (RE)}\label{sec:re}
The radio environment module reads the signal strength streams, and matches the observed 
signal changes to a specific workstation. A user who steps away from the workstation 
alters signal propagation of wireless sensors, based on trajectory of movement.
These signal alterations form a recognizable pattern and we use a classifier to identify them.
We now describe how training samples are built and labeled, as well as how the system is trained.

\subsubsection{Samples}
Each sample represents a signature of the effect on the radio environment of
a user who after sitting at a workstation leaves the proximity of that workstation. 

To obtain a sample, we need to identify the correct interval where the user leaves the 
proximity of the workstation. We use the timings provided by the variation windows 
measured by MD module, assuming that movement is always obtained with a variation 
window $[t_1, t_2]$. When such a window is observed, for each stream $i$, we extract 
signal strengths observed in window:
$$V_{t_1, t_1+t_\Delta}^{(i)} = \{r^{(i)}_{t_1}, r^{(i)}_{t_1+1} ..., r^{(i)}_{t_1+t_\Delta}\}$$
of size $|V_{t_1, t_1+t_\Delta}^{(i)}| = n$, and we compute the following features:
\begin{compactitem}
\item \textbf{Variance} of the window: 
\[\sigma^{2} = \frac{\sum\limits_{j}^{} (r_j^{} - \mu^{})^2}{n}. \]
\item \textbf{Entropy} of the frequency distribution histogram $V$ of the window:
\[H^{} = - \sum\limits_{r_j \in V}^{} P(r_j) \log P(r_j). \]
\item \textbf{Autocorrelation} of the window:
\[ R(k) = \frac{1}{(n-k)\sigma^{2}} \sum\limits_{j=t_1}^{n-k}(r_j-\mu)(r_{j+k} - \mu).\]
\end{compactitem}
We use the window from $[t_1, t_1 + t_\Delta]$ instead of the full $[t_1, t_2]$ because the most 
distinctive part of the user's physical position when leaving the workstation occurs at the beginning. 
Some portions of the user's path out of the office are likely to overlap with 
paths of other users (since there is a single door, users likely move towards it).
Meanwhile, initial segments of users' paths are naturally less likely to overlap.

\subsubsection{Labels}
We use a set of labels $w_0,\:w_1,\: ...,\: w_k$ for samples associated with specific events.
$w_0$ is associated with the event: ``user entered the office'' and each other $w_i$ 
is associated with the event: ``user left workstation $w_i$''.  $w_0$ is needed since
users entering the office generate significant fluctuations in the radio environment: this
must be detected so as not to mistakenly deauthenticate users who are still at their workstations.

\subsubsection{Training Phase}\label{sec:training_phase}
During the training phase, \acro\ runs the MD module to detect variation windows.
For every variation window, \acro\ extracts the corresponding sample, computing its features.
Afterwards, the system uses KMA to fetch idle time for the workstations, and tries to automatically 
label the sample. The time of a user's departure from the workstation and idle time at that workstation 
are correlated, since from the moment the user walks away the workstation observes no further
input, until the user returns. To avoid erroneous labeling, when \acro\ is uncertain (i.e., more than one 
workstation is idle during the variation window) it simply discards the sample.
At the end of the training phase, labeled samples are used to set up 
a Support Vector Machine (SVM) classifier used in the online phase.

\subsubsection{Online Phase}
In this phase, whenever MD observes a variation window $W = [t_1, t_2]$ of size 
$\geq t_\Delta$, at time $t_1 + t_{\Delta}$ (before the end of the window) the system 
queries RE. The module fetches the window of signal strength measurements 
$V^{(i)}_{t_1, t_1+t_\Delta},  \forall \: i$, computes the features for the windows 
to construct the sample, inputs the sample into the classifier and returns the predicted label.

\subsection{Overlaps}\label{sec:overlap}
\begin{figure}[htpb]
    \centering
    \includegraphics[width=0.45\textwidth]{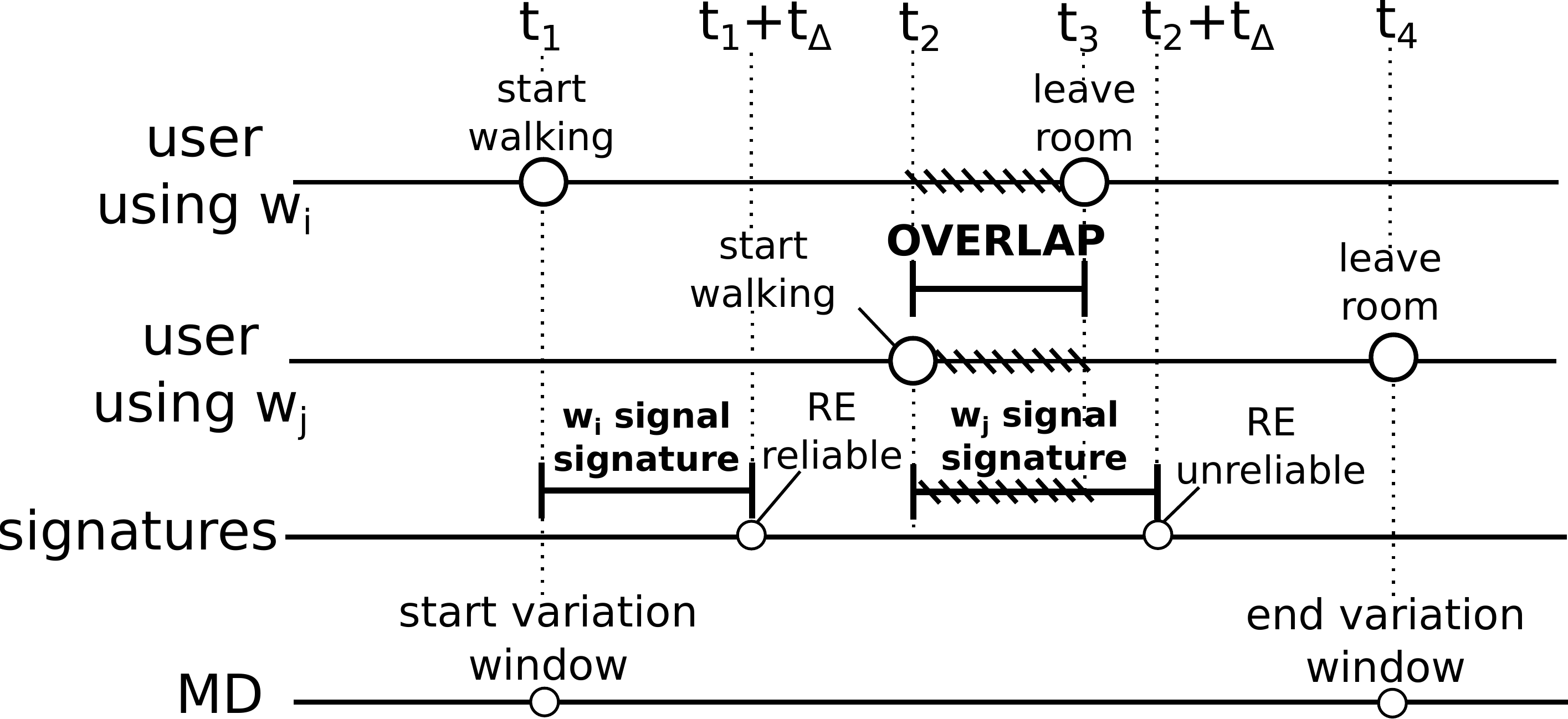}
    \caption{Sample overlap timeline.}\label{fig:overlaps}
\end{figure}

\acro\  does not recognize situations when multiple users walk away from their workstation at the same time,
or their movement intervals overlap. RE is trained with samples that correspond to events where, from an
initial state when no one is moving away,  a single user leaves the workstation. Figure~\ref{fig:overlaps} 
shows an example where the movement of one user interferes with other users signatures and vice-versa.
Two users at $w_i$ and $w_j$ walk away and MD observes a single variation window $[t_1, t_4]$; note that
both users are moving inside the room in the interval $[t_2,t_3]$. 
We refer to this situation as \textit{overlap}. Whenever it happens, signatures in the radio environment 
are unreliable, since multiple bodies in motion alter radio signals in different physical locations.
Moreover, since MD only detects discrepancies from a normal fluctuations profile, \acro\ can not 
detect whether multiple users are moving.

To tackle this issue, \acro\ errs on the conservative side: as long as MD observes 
the continuation of the variation window after $t_1+t_\Delta$, \acro\ accounts for the possibility 
of other users possibly leaving and triggers activity responses at the workstations based on 
their idle time.  The next section details activity responses and their trigger events.

\subsection{System Actions and Rules}\label{sec:rules}
To introduce the final system flow, we identify the types of actions \acro\ can impose 
on the workstations:
\begin{compactitem}
\item \textbf{Deauthenticate:} current login session on $w_i$ is deauthenticated, 
and requires re-authentication.
\item \textbf{Alert State:} in this state, if $w_i$ remains idle for at least $t_{ID}$ seconds, the system 
activates a screen saver. However, if any keyboard or mouse activity is observed, $w_i$ exits 
alert state and current session remains authenticated.
The duration of alert state is only few seconds, as discussed in Section~\ref{sec:experimental_results}.
\end{compactitem}
Table~\ref{tab:rules} shows the rules that \acro\ uses to determine the appropriate action. 
Columns \textbf{RE} and \textbf{KMA} show the outputs from the modules, 
while column \textbf{Action} describes the type of action. 

\begin{table}[htpb]
\begin{center}
\begin{tabular}{c|c|c|c}
   \textbf{Rule} & \textbf{RE} & \textbf{KMA} & \textbf{Action} \\
 \hline \textbf{1} &  $c_i$ &  $S^{(t_{\Delta})}$ & if $c_i \notin S^{(t_{\Delta})}$ then Deauthenticate $c_i$ \\ 
 \hline \textbf{2} & - & $S^{(1)}$ & $\forall \: c_i \in S^{(1)} \text{ Alert State } c_i$ 
\end{tabular} 
\caption{Rules used to determine the action. }
\label{tab:rules}

\end{center}
\end{table}

\subsection{System Workflow}\label{sec:system_workflow}
The system works as a finite state automaton with two states; in each discrete 
time step it queries MD, and, depending on its output, moves between the states.  State
transitions are defined based on the duration of the current variation window reported by MD. 
At time $t$ we refer to $W_t = [t_i,t]$ as the most recent variation window, lasting from
$t_i$ until $t$. If such a variation window does not exist, we assume that $t-t_i=0$.
Let $d_{W_t} = t - t_i$ be the size of this window.

\begin{figure}[htpb]
    \centering
    \includegraphics[width=0.45\textwidth]{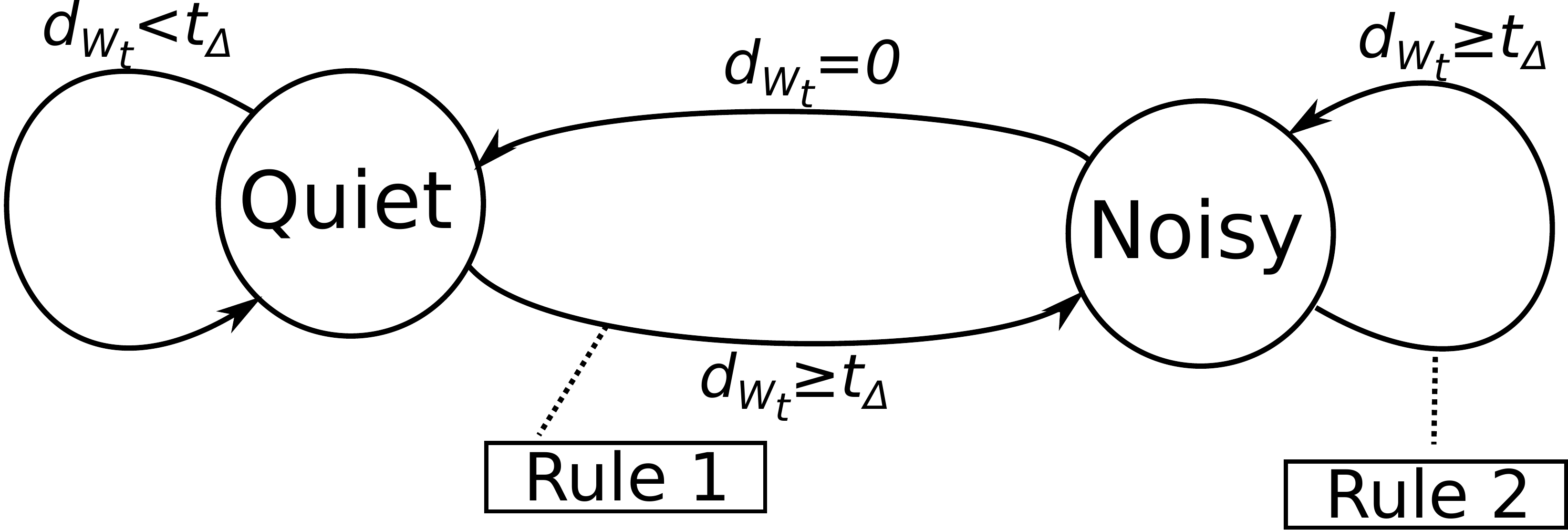}
    \caption{\acro\ state diagram.}\label{fig:methodology-flow}
\end{figure}

Figure~\ref{fig:methodology-flow} shows the state diagram. 
Two states: \textbf{Quiet} and \textbf{Noisy} allow \acro\ to react depending on the 
conditions of the radio environment. \textbf{Noisy} deals with overlaps mentioned in 
Section~\ref{sec:overlap}, while \textbf{Quiet} handles normal cases when 
users leave their workstations one at a time. The system remains in \textbf{Quiet}, until 
MD observes variations window of less than $t_{\Delta}$, i.e., $d_{W_t} < t_{\Delta}$. 
As soon as the current variation window reaches $t_{\Delta}$ (i.e., $d_{W_t} = t_{\Delta}$), 
the system queries RE and KMA, and applies Rule 1. Next, state transitions to \textbf{Noisy}.
While in it, until the current variation window continues (i.e., $d_{W_t} > t_{\Delta}$), 
at each time step the system queries KMA and applies Rule 2.
When MD reports that the variation window is over (i.e., $d_{W_t} = 0$), \acro\ 
transitions back to \textbf{Quiet} and repeats the cycle.

\section{Security Analysis}\label{sec:security_analysis}
We now analyze security issues, focusing on potential attacks. 
Section~\ref{sec:assump_and_term} discusses possible outcomes and
\mbox{Section~\ref{sec:sec_modeling}} considers their security implications.

\subsection{Terminology}\label{sec:assump_and_term}
We first categorize decisions by MD and RE.
For MD, when the user leaves the workstation at time $t$, we consider
the interval where a movement should be observed by MD. We refer to it as the \textit{true} window: 
$U_{t}=[t-\delta,t+\delta]$. Given all  true windows $ \{U_{t_i}, ..., U_{t_j}\}$, and all variation windows 
$ \{W_{t_h}, ..., W_{t_k}\}$ observed by MD, we classify MD decisions as:
\begin{compactitem}
\item \textbf{True Positive }(TP): $W_{t_h}$ and $U_{t_i}$ overlap -- correctly identified 
movement.
\item \textbf{False Positive }(FP): $W_{t_h}$ does not overlap with any $U_{t_i}$ -- incorrectly 
identified movement.
\item \textbf{False Negative }(FN): $U_{t_i}$ does not overlap with any $W_{t_h}$ -- failure to identify 
movement.
\end{compactitem}
For RE, we only consider true positives. False positives do not represent an attack opportunity 
since no workstation is left unattended (they affect usability), while false negatives imply that 
MD did not observe the variation window; therefore \acro\ does not interrogate RE, as 
discussed in Section~\ref{sec:system_workflow}.
When a user leaves $w_i$ leaves and a true positive occurs, \acro\ queries RE to classify 
the sample corresponding to the observed variation window. There are two possible outcomes:
\begin{compactitem}
\item \textbf{Correct}: RE outputs $w_i$,
\item \textbf{Mis-classified}: RE outputs $w_h \neq w_i$.
\end{compactitem}
As a baseline, we assume that the workstation has a normal deauthentication time-out, i.e.,
whenever it is idle for $T$ seconds, the current session is terminated.
We also assume that last input of a user departing at time $t$ occurs at
exactly that time. This is a worst-case assumption: if last input occurrs earlier, 
deauthentication takes place sooner.  In the next section we fuse these considerations 
to obtain a comprehensive model for security analysis.

\subsection{Security Modeling}\label{sec:sec_modeling}
\begin{figure}[htpb]
    \centering
    \includegraphics[width=0.45\textwidth]{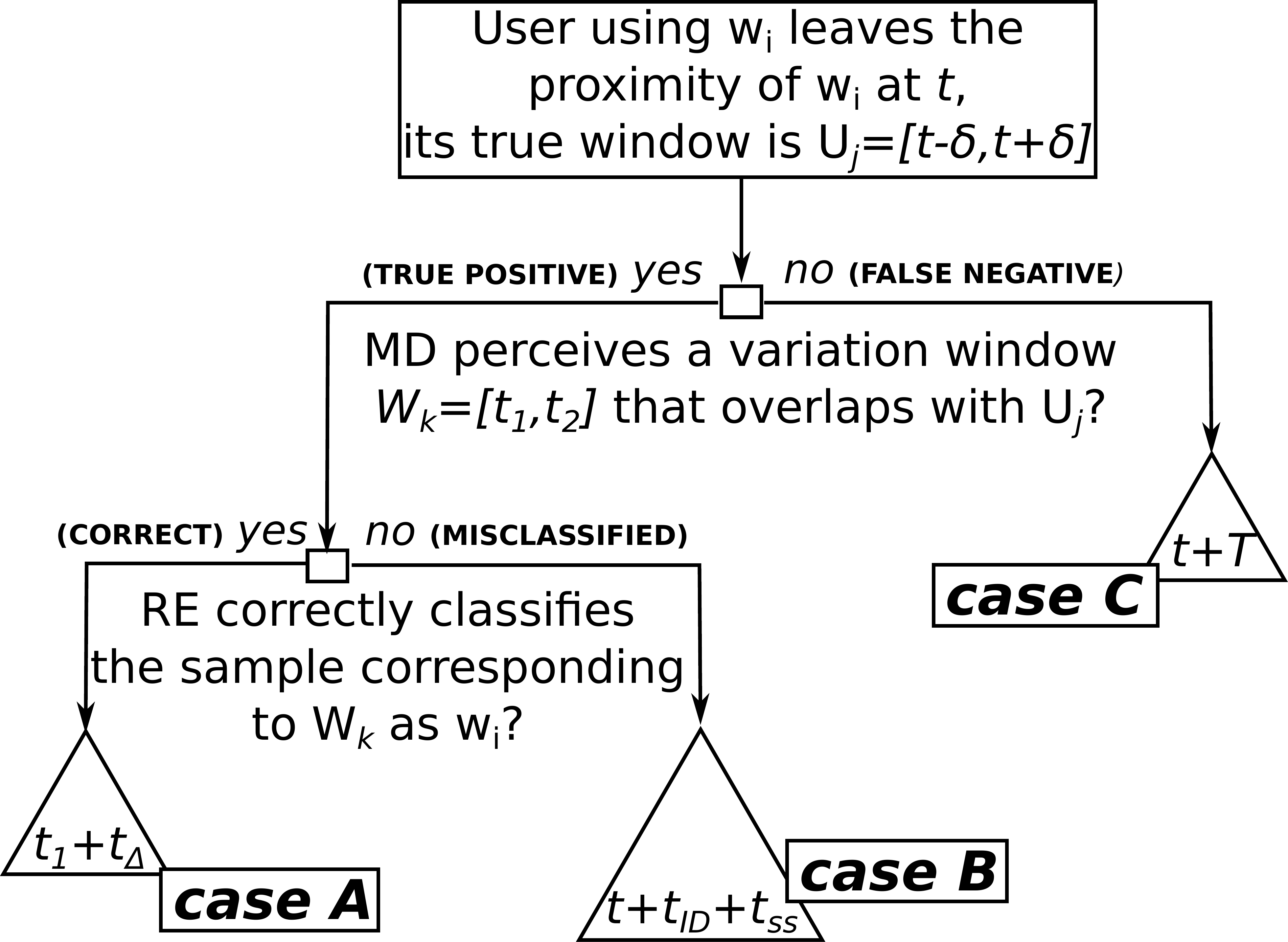}
    \caption{Decision tree showing deauthentication timings.}\label{fig:dec_tree_sec}
\end{figure}

We use decision tree analysis to create a tree that represents all possible outcomes 
when a user leaves the workstation.  Figure~\ref{fig:dec_tree_sec} is a representation of the 
tree; its leaves contain the times when deauthentication occurs, for each case. It shows that true 
positives, depending on RE outcome, may result in deauthentication at time $t_1+t_\Delta$ 
when the classification is correct (case A), or deauthentication at time $t+t_{ID}+t_{ss}$ 
when the sample is misclassified (case B). False negatives result in deauthentication 
by time-out at time $t+T$ (case C). 

\subsection{Wireless Physical Attacks}\label{sec:resilience}
One intuitive way to attack \acro\ is by jamming, which alters wireless signal propagation 
and RSSI. Feasibility of attacks that manipulate wireless signals  in real-time has  been
demonstrated in real-world scenarios: signal can be either annihilated entirely, or its 
content can be modified~\cite{popper2011investigation}. In scenarios where RSSI integrity 
is critical (such as WLAN-based positioning systems), jamming attacks can seriously 
compromise the outcomes~\cite{tippenhauer2009attacks}. 

In our case, we believe that the adversary can not jam the wireless signal to alter RSSI.
To prevent correct operation of \acro, the adversary must alter all wireless transmission 
among devices such that standard deviations of signal streams do not increase when 
a users steps away from the workstation.
To do so, the adversary must determine the current RSSI of a stream and alter it to 
make sure that new RSSI measurement matches the previous ones. Since RSSI strongly 
depends on locations of communicating device and impact of 
physical obstacles (i.e., users and other objects in the room), we believe that the adversary 
can not alter signal strength obtained by specific sensors at specific times, i.e., when the 
moving user alters signal strengths between specific sensors.

Even if such alterations were achievable, due to close relative proximity of devices, it is 
very hard to limit alteration only to certain devices, i.e, the alteration of one transmission 
originating from device $d_i$ is measured by all the other devices. Therefore, such attacks 
are detectable. Thus, we believe that such physical attacks are ineffective against \acro.

\section{Experimental Design}\label{sec:experimental_design}
In this section we discuss design choices for the experiments. 

\subsection{Design Choices}\label{sec:design_aims}
The goal of the experiments is to show that \acro\ is both secure and usable. 
To assess security, we use the time required for  deauthentication, as described in 
Section~\ref{sec:sec_modeling}. This is reasonable in our model, where the adversary can 
can physically access an unattended workstation with an active (authenticated) login session.
To measure this time, it is sufficient to measure frequencies of occurrence for each leaf in the 
decision tree of Figure~\ref{fig:dec_tree_sec}. We record the timing when the user exits
the office.  This allows for realistic measurements of the impact of two adversary types (Insider and 
Co-worker), in terms of number of their opportunities to attack without being witnessed.

For the usability aspect, we account for the fact that the system may incorrectly activate a  
screen saver or perform deauthentication, when the user is still present at the workstation. 
In such cases, the user needs react by either canceling the screen saver or re-authenticating.
This extra effort (cost) can be viewed as a fixed delay. Screen saver cancellation and deauthentication
involve different costs, since the the former only requires the user to generate some input, 
while the latter involves re-authentication, i.e., a new log-in. 

To evaluate \acro's performance in a general setting we need to make as few assumptions as possible.
To account for differences in keyboard typing or mouse movement habits among users, we 
simulate the keyboard and mouse input on the workstations.
This is because many system decisions depend on idle time observed at the workstations, 
and we want to avoid non-representative behavior of our subjects compromise the evaluation.

\subsection{Experiment Structure}\label{sec:exp_structure}
\begin{figure}[htpb]
    \centering
    \includegraphics[width=0.45\textwidth]{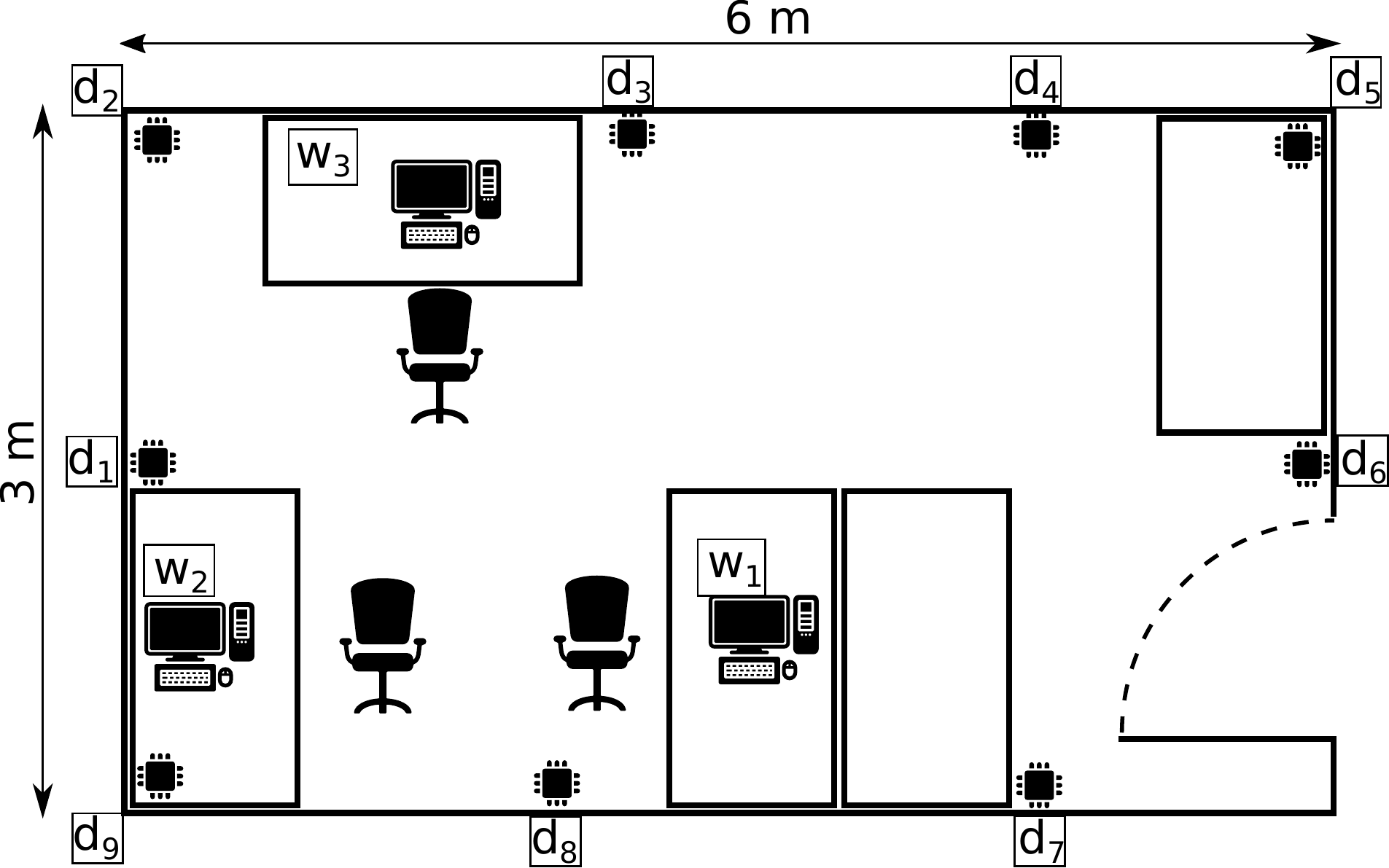}
    \caption{Layout of the office where experiments were performed.}
    \label{fig:experiment_setup}

\end{figure}

We conducted experiments in one of our offices, that adheres to our system model assumptions. 
The office contains three workstations and three distinct users (students), each assigned to
exactly one workstation. We placed nine wireless sensors along the office walls, about one meter 
from the ground; slightly above the average desk height. Figure~\ref{fig:experiment_setup} shows the 
layout of the office as well as sensor and workstation locations.
Users were not required to perform any extra tasks during the experiments, in order to emulate 
their everyday routine. A human supervisor monitored actual timings of user movements by noting the times when 
users stepped away from their workstations, as well as the times when they entered and exited the room.
We collected signal strengths observed by the sensors for five consecutive days, during working hours 
(9am-5pm), for a total of 40 hours. At the end, we obtained $130$ labeled events, summarized in 
Table~\ref{tab:data}. We did not register any overlap in the collected data; see Section~\ref{sec:overlap}. 

\begin{table}[htpb]
\begin{center}
\begin{tabular}{c|c|c|c|c}
   \textbf{label} & \textbf{$w_0$} & \textbf{$w_1$} & \textbf{$w_2$} & \textbf{$w_3$} \\
   \hline  
  \textbf{number of events} & \textbf{$67$} & \textbf{$21$} & \textbf{$20$} & \textbf{$22$} \\
\end{tabular} 
\caption{Number of labeled events obtained during data collection.}
\label{tab:data}
\end{center}
\end{table}

\section{Experimental Results}~\label{sec:experimental_results}
We now present experimental results and analyze \acro performance improvement 
when the number of sensors is increased.

\subsection{MD performance}\label{sec:md_performance}
\begin{figure}[htpb]
    \centering
    \includegraphics[width=0.45\textwidth]{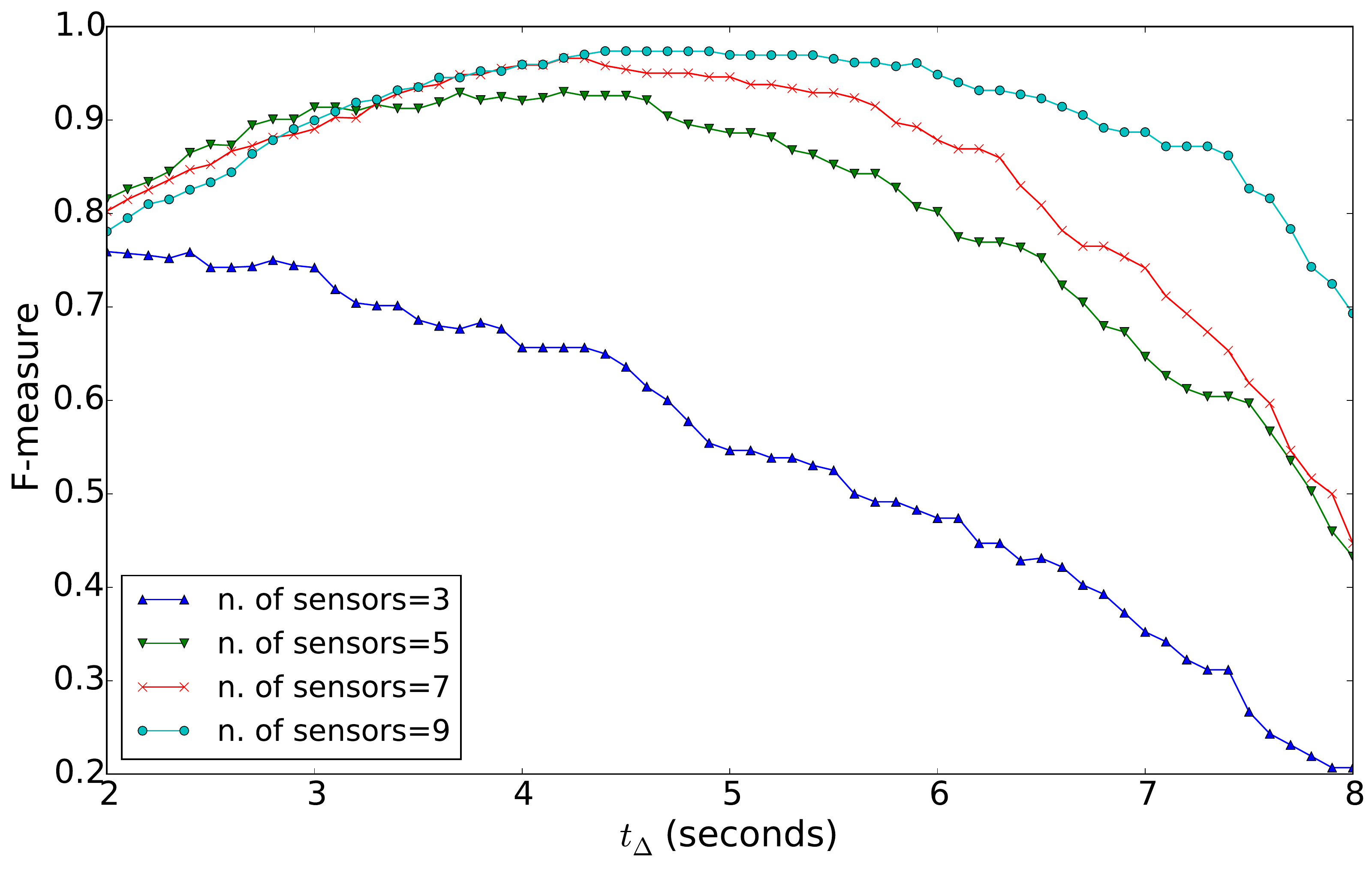}
    \caption{F-measure for MD, for varying values of $t_\Delta$.}\label{fig:md_performance}

\end{figure}

We measure MD performance in terms of TP, FP, and FN. 
Figure~\ref{fig:md_performance} shows the F-measure for MD computed as 
$2\cdot \frac{precision\cdot{}recall}{precision+recall}$, for increasing sizes of parameter $t_\Delta$, 
and for different number of sensors.
The F-measure shows the trade-off between the number of variation windows correctly identified as TP, and
the number (out of the 130 events) of correctly identified events, by variation windows. 
Figure~\ref{fig:md_performance} shows that there is an F-measure peak around $t_\Delta=5.0$.
This is expected, since in the experiment room the average time needed by a user to walk from the 
workstation to the door is about 5 seconds (4-meter distance, assuming walking speed of 1.4 m/sec, 
plus time required to stand up and later open the door). 
Recall is more important than precision, because for each FN case C happens 
(see Figure~\ref{fig:dec_tree_sec}). Therefore, we err on the side of caution and 
use $t_\Delta=4.5$; hereafter, all results refer to this value., unless otherwise specified.
Table~\ref{tab:md_tables} shows percentages of obtained TP, FP and FN for various 
numbers of sensors for $t_\Delta=4.5$.
The table shows a promising result: increasing the number of sensors 
becomes quickly conservative against FN, recording zero of them with 8 or more sensors.

\begin{table}[htpb]
\begin{center}
\begin{tabular}{c|c|c|c}
   \textbf{n. of sensors} & \textbf{TP} (\#) & \textbf{FP} (\#)& \textbf{FN} (\#) \\
\hline \textbf{3} & 0.47 (62) & 0.02 (3) & 0.51 (68) \\
\hline \textbf{4} & 0.77 (106) & 0.05 (7) & 0.18 (24) \\
\hline \textbf{5} & 0.86 (119) & 0.06 (8) & 0.08 (11) \\
\hline \textbf{6} & 0.88 (122) & 0.06 (8) & 0.06 (8) \\
\hline \textbf{7} & 0.91 (125) & 0.05 (7) & 0.04 (5) \\
\hline \textbf{8} & 0.96 (130) & 0.04 (6) & 0.00 (0) \\
\hline \textbf{9} & 0.95 (130) & 0.05 (7) & 0.00 (0) \\
\end{tabular} 
\caption{MD performance in percentage in terms of true positives, false positives and false negatives data.}
\label{tab:md_tables}

\end{center}
\end{table}

\subsection{RE performance}\label{sec:re_performance}

\begin{figure}[htpb]
    \centering
    \includegraphics[width=0.45\textwidth]{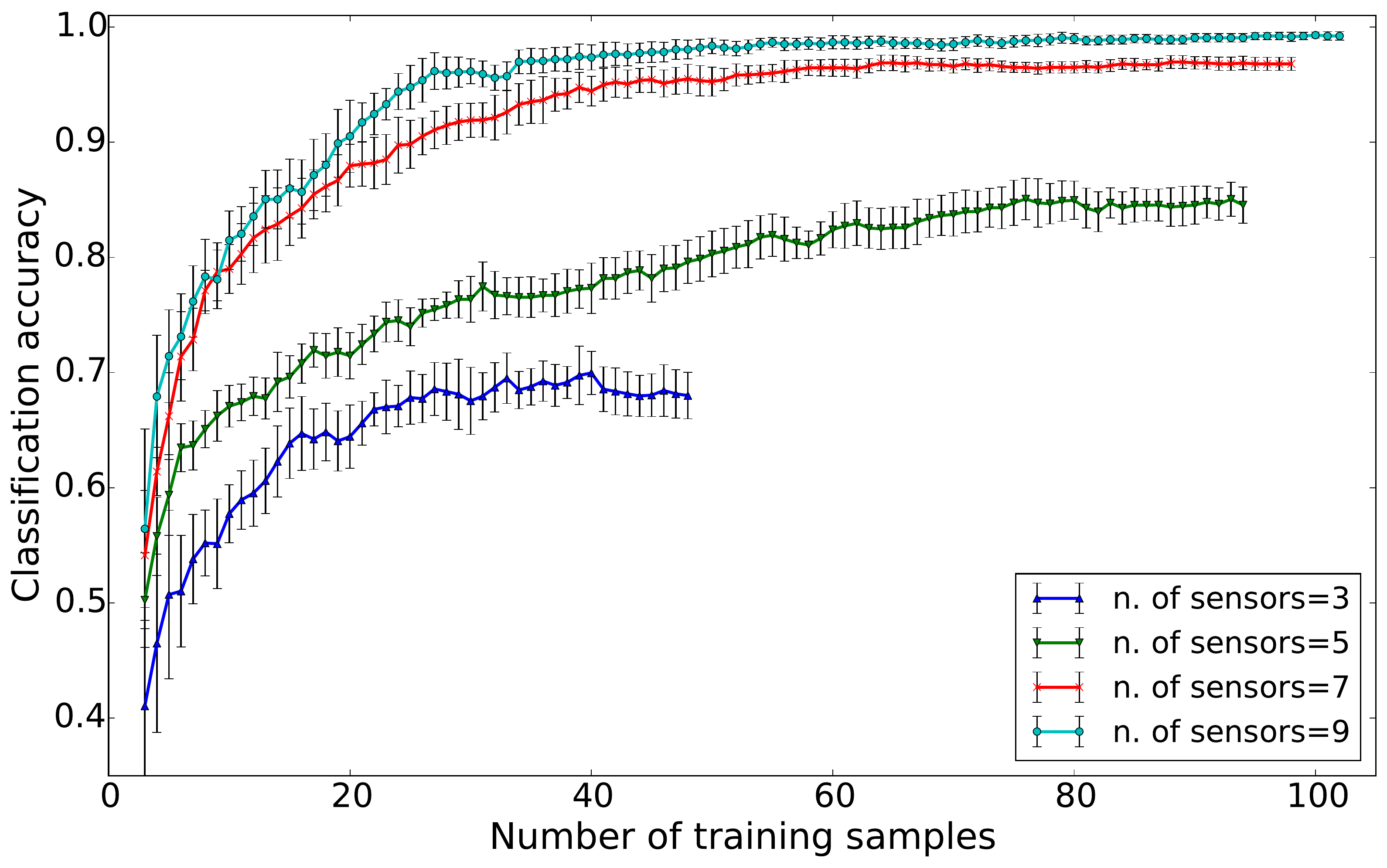}
    \caption{Accuracy of the classification for RE, for an increasing number of samples in the training set.}\label{fig:re_accuracy}

\end{figure}

We measure the performance of RE in terms of the accuracy of the classification of the variation windows that correspond to a TP. 
In order to measure it, we split the collected data into training and test set in a 5-fold validation. 
For each fold, we train the classifier with increasing number of samples in the of training set, and compute the accuracy on the test set for each size.
Since we have a relatively small number of samples, we repeat the process 10 times to account for the differences in the cross validation random split.
Figure~\ref{fig:re_accuracy} shows the accuracy of the classification for an increasing number of training samples, averaged over the 5 fold of the validation.
The error bars show the 95\% confidence interval on the 10 different splits for the cross validation.
Since considering fewer sensors some events have resulted in a smaller number of TP (see Table~\ref{tab:md_tables}), some of the lines end early on the x-axis. 
Figure~\ref{fig:re_accuracy} shows that a classifier is able to learn quickly how to discriminate between different workstations. 
In fact, for 7 or more sensors, after only 40 samples (that correspond roughly to 2 days of training), RE reaches an accuracy greater than 90\%.
The figure also shows that the more sensors are available and the steeper is the learning curve, and the more accurate the classification becomes.

\subsection{Security}\label{sec:eval_security}
As mentioned in Section~\ref{sec:design_aims}, our performance indicator for the security is the time required for the deauthentication of the workstation after a user left its proximity. 
To measure this time, and hereafter in this section, we run the system on our data as follows: first we run MD on the whole monitored period, and obtain its TP, FP, and FN, then we split the obtained samples into a 5-fold validation. 
For each fold, we train RE with the TP in the training set, and for each TP in the test set we classify it with RE.
Given the output of RE and MD, we check in which of the cases illustrated in Figure~\ref{fig:dec_tree_sec} we end up.
For each case, we compute the time required for the deauthentication accordingly.

\begin{figure}[t]
    \centering
    \includegraphics[width=0.45\textwidth]{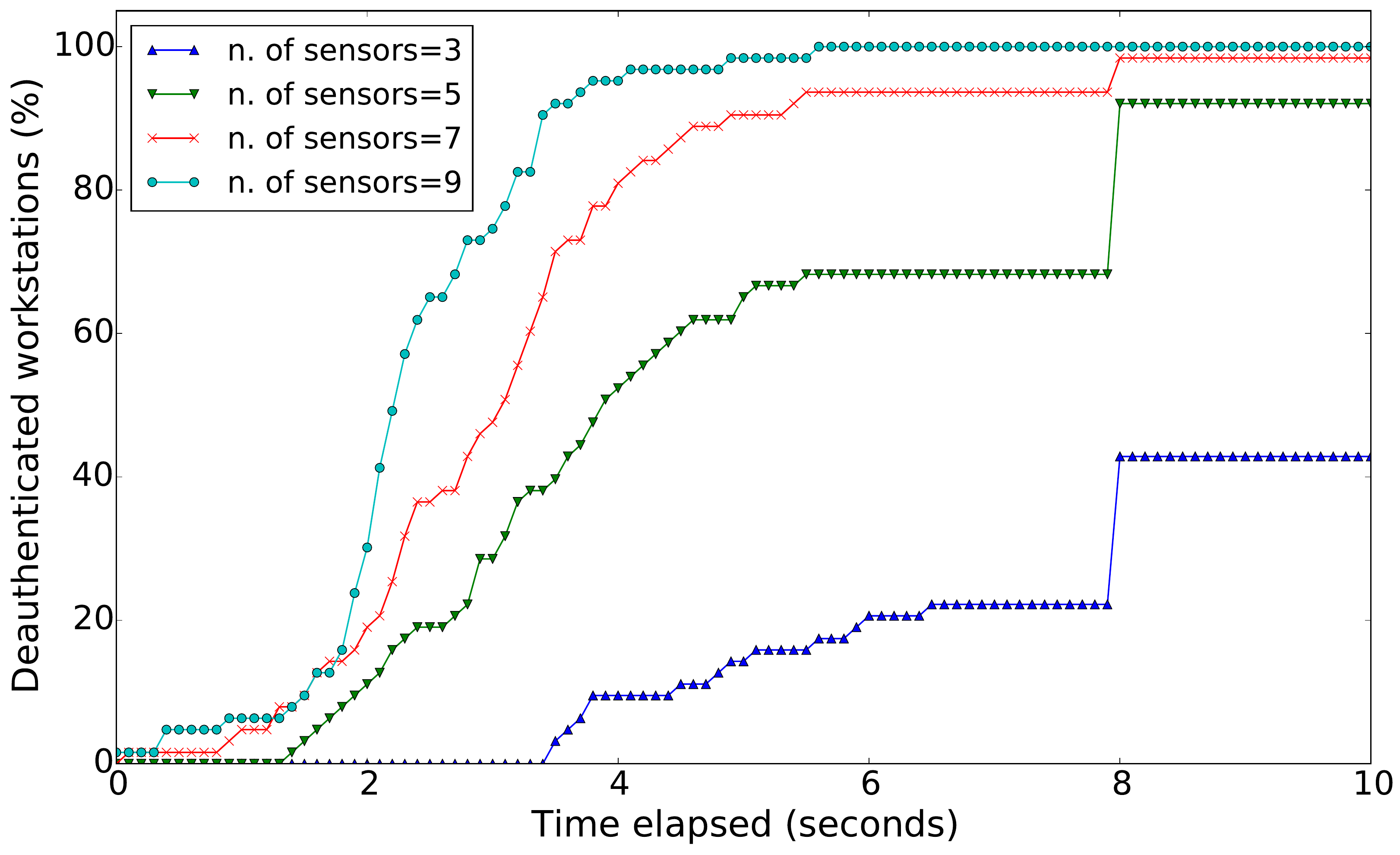}
    \caption{Proportion of deauthenticated workstations.}\label{fig:secured_ws}
    
\end{figure}

Figure~\ref{fig:secured_ws} shows the proportion of deauthenticated workstations for increasing time elapsed after the user left, for $t_\Delta=4.5$, $t_{ID}=5$ and $t_{ss}=3$. 
As shown in the plot, increasing the number of sensors leads to a faster deauthentication, and a greater number of events that are captured by MD (case A and case B). 
There is a step in the curves exactly at 8 seconds, this shows the amount of events that have been misclassified by RE, resulting in case B, where the deauthentication occurs always after $t_{ID} + t_{ss}=8$ (as shown in Section~\ref{sec:rules}) from the last user input.
The occurrence of case C leads to some workstations not being deauthenticated after 10 seconds, these workstations are deauthenticated after the expiration of the baseline time-out $T$.

\begin{figure}[b]
    \centering
    \includegraphics[width=0.45\textwidth]{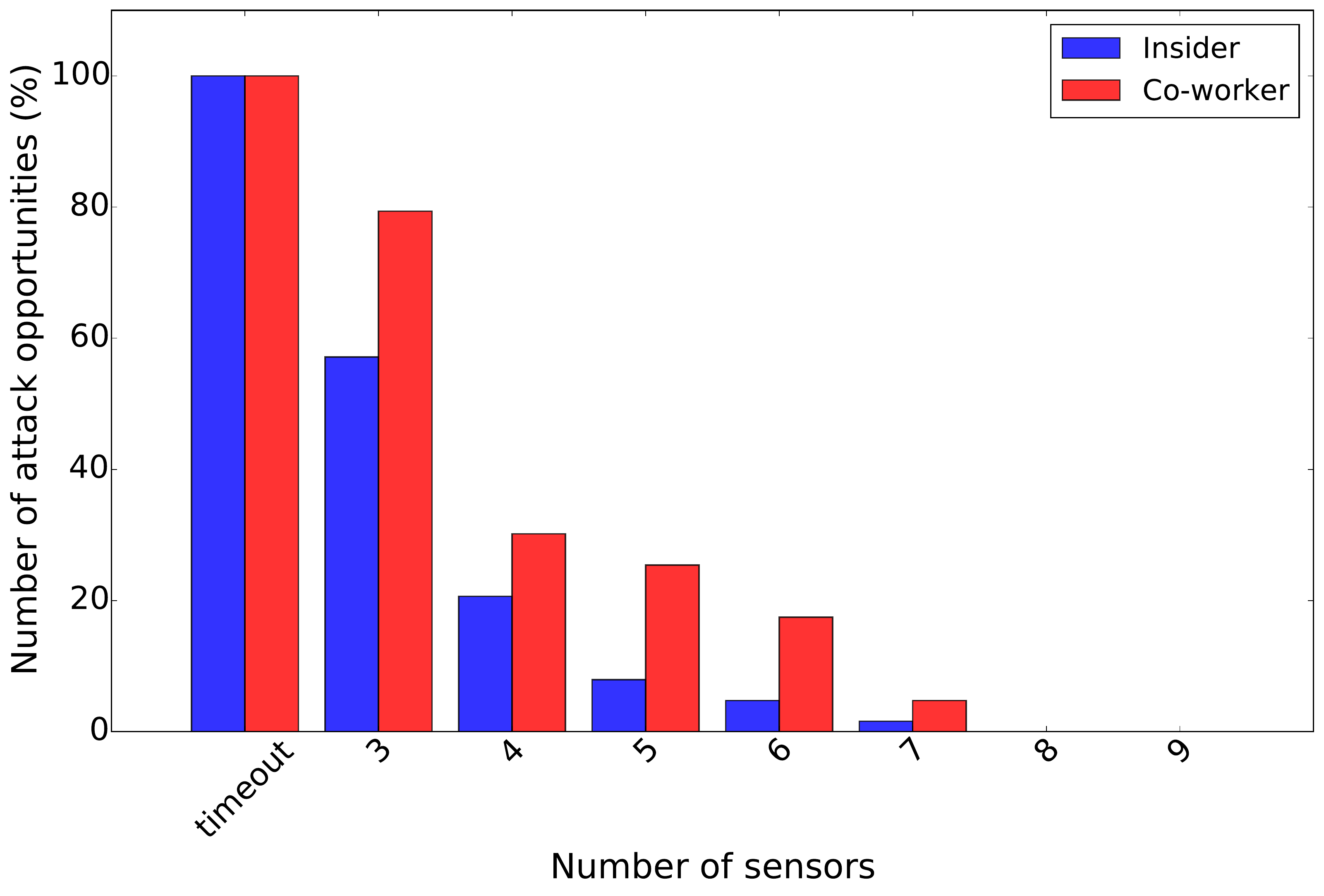}
    \caption{Percentage of times adversaries can perform an attack on a target workstation, for an increasing number of sensors.}\label{fig:attackers}

\end{figure}

Figure~\ref{fig:attackers} shows the number of opportunities that adversaries have to access the target workstation without being witnessed by the victim, for Insider and Co-worker, respectively.
For Insider we consider that he can reach the workstation after 4 seconds since the victim left the office (this a realistic estimate of the time required to walk to the workstation from outside the office).
For Co-worker, we consider that he can reach the workstation as soon as the victim user left the office.
We consider that every time a user leaves his workstation unattended, an attack is possible.
As shown in the plot, with an approach with time-out, both adversaries can perform an attack every time a user leaves the office (63 times in our experiment).
Increasing the number of sensors, the attack opportunities significantly decrease, down to zero for 8 or more sensors.

\subsection{Usability}\label{sec:eval_usability}

To simulate the keyboard and mouse input at the workstations, we refer to the work of Mikkelsen et al~\cite{Mikkelsen2007}, where they monitored the keyboard and mouse usage of a sample of 1211 office workplace users.
In their analysis, they discretized time in 5 seconds intervals, and found that, on average, users are using the keyboard or the mouse during 78\% of these intervals.
We refer to their findings, and we simulate the user inputs in each 5 seconds intervals, such that users have a 78\% probability of using the mouse or keyboard during an interval.

We assign a cost (as defined in Section~\ref{sec:design_aims}) of $3$ seconds for deactivating a screen saver (some users just remove it before its expiration), and a cost of 13 seconds to re-perform the authentication (this is on average the time users need to correctly input their login information~\cite{Shay2014}).
Moreover, we assume that when a user leaves the proximity of his workstation, all the other users are inside the room and they are using their own workstation (this is a worst case assumption, since some of them might not be in the office and therefore not be impacted by the system decisions).

For the simulation we follow the procedure described in Section~\ref{sec:eval_security}, plus we randomly draw the users keyboard and mouse input distribution over the monitored period.
Since the outcome of the system is dependent on the user inputs, we draw this distribution for 100 times to account for possible differences, and average the result.

Table~\ref{tab:usability} shows the total cost for different number of sensors, computed by multiplying the average number of screen savers and deauthentications for their respective cost (i.e, 3 and 13 seconds, respectively).
Values between parenthesis show the standard deviation over the 100 runs of the simulation.
The results show that while the number of deauthentication decreases with more sensors, due to the improved precision of RE, the number of screen savers does not decrease.
This happens because with more sensors we have an higher recall for MD, and therefore the system needs to deal with an increased number of variation windows, and activates more screen savers.
However, Table~\ref{tab:usability} shows that the total cost each day is never higher than 37 seconds. 
This means that on average, one user (we have three users in total) is required to invest ${\sim}13$ seconds each day to collaborate with the system, which is a very acceptable time.

\begin{table}[htpb]
\begin{center}
\begin{tabular}{c|c|c|c}
   \specialcell[c]{\textbf{n. of} \\ \textbf{sensors}} & \specialcell[c]{\textbf{\# screen savers} \\ \textbf{per day}} & \specialcell[c]{\textbf{\# deauthentication} \\ \textbf{per day}} & \specialcell[c]{\textbf{cost (seconds)} \\ \textbf{per day}} \\
\hline \textbf{3} & 4.272 (0.82) & 0.712 (0.36) & 22.07 \\
\hline \textbf{4} & 8.238 (1.02) & 0.926 (0.43) & 36.75 \\
\hline \textbf{5} & 8.336 (1.09) & 0.754 (0.38) & 34.81 \\
\hline \textbf{6} & 8.414 (1.17) & 0.558 (0.29) & 32.5 \\
\hline \textbf{7} & 8.188 (1.03) & 0.136 (0.15) & 26.33 \\
\hline \textbf{8} & 8.956 (1.22) & 0.086 (0.12) & 27.99 \\
\hline \textbf{9} & 9.094 (1.15) & 0.036 (0.09) & 27.75 \\
\end{tabular} 
\caption{Number of times the system takes an incorrect decision, and total cost in seconds, for a 8h period.}
\label{tab:usability}

\end{center}
\end{table}

\section{Conclusions}\label{sec:limitations_and_future_works}
Current research on authentication mechanism focuses more on the authentication part, and does not thoroughly take into account the importance of the deauthentication procedure.
Even if authentication solutions that provide automatic deauthentication exist, they have limitations: they require the user to carry additional devices, they lack realistic threat models, or they require expensive hardware.
In this work, we proposed a solution for the deauthentication -- \acro\ -- that leverages physical properties of wireless signal propagation, in order to secure unattended workstations in the office workplace.
\acro\ uses the information provided by wireless sensors placed inside the office to learn the behavior of the radio environment, and deauthenticate users when they leave the vicinity of their workstations.
In order to evaluate the security and usability of our system, we designed an experiment with very few assumptions, and we carried it out in one office.
We showed that, even in a small office where the radio environment is dynamic and busy, our system grants good performance.
In particular, when the sensors grant a sufficient coverage of the area inside the office, the system becomes very accurate and fast in the deauthentication (adversaries cannot exploit any opportunity to perform an attack), without disregarding the usability.

\subsection{Discussion and Future Work}
The research question behind this paper, was to understand if it was possible to detect and learn the changes in the wireless propagation caused by the movement of users.
However, characterizing the behavior of wireless signals in such dynamic, small and cluttered environments has proved to be challenging.
Even if the physical phenomena that regulate the propagation of radio signals are known, we could not use them directly to model the environment in our case.
In order to obtain an effective system, we had to rely on high level observations on the monitored signal strengths, use machine learning for the classification, and combine these information with the user inputs at the workstation.
Even if this is a poor modeling of the movement of users, we realized that in a system with such weak and loose assumptions (e.g., busy wireless channel, multiple users offices, cheap wireless sensors with simple hardware), our results are noteworthy.

In the future, in order to prove that the system can be effective in different environments, we plan to investigate the performance of the system in different setups (other offices, with different dimensions and users).
We also want to evaluate its performance considering different placements of the sensors, to understand if the wireless devices currently present in a common office (e.g., desktop computers, Internet of Things devices) are sufficient to obtain valuable results.
Furthermore, we also want to explore whether more fine grained information that can be provided by the wireless channel (such as channel state information) can improve the system performance.

\clearpage
\balance
\renewcommand*{\bibfont}{\small}
\printbibliography
\clearpage
\appendix
\nobalance

\subsection{Feature Analysis}

\begin{figure}[htpb]
    \centering
    \includegraphics[width=0.45\textwidth]{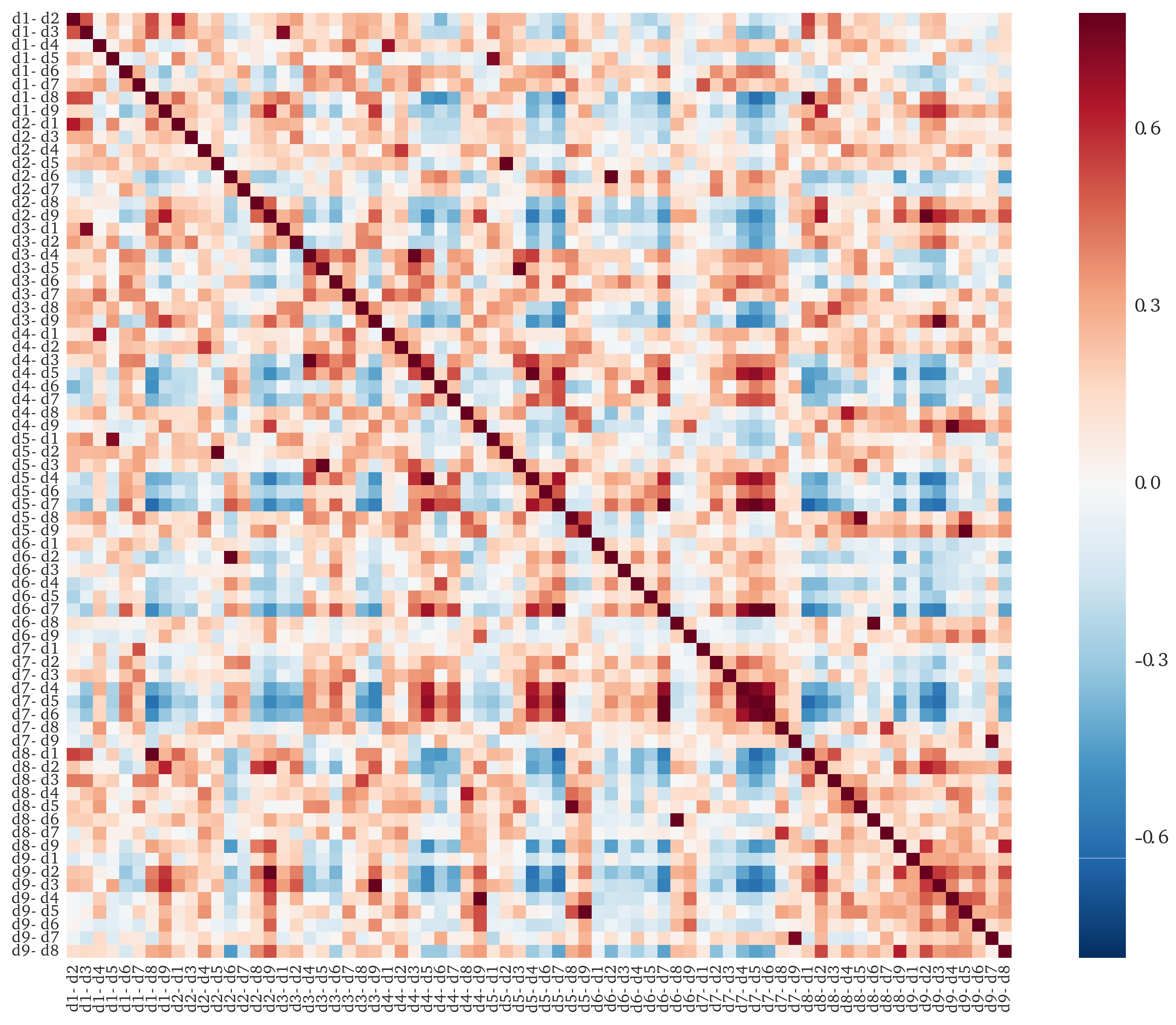}
    \caption{Correlations between the variances of streams in the collected samples.}\label{fig:correlations}
\end{figure}

To understand the role of the features of the samples for RE, we study the correlations between them, and their importance in the classification, using all of the nine sensors.
Figure~\ref{fig:correlations} shows the correlation between the variances of all the streams, computed over the labeled samples (for brevity, we did not include the entropies and the autocorrelations).
In the figure, the label $d_i-d_j$ identifies the variance of the stream that goes from device $i$ to device $j$.
As shown, when the devices are close to each other, their variance reacts in similar ways to the alteration caused by the user who is moving.

In order to measure the importance of the features, we remove highly correlated and uncorrelated features, and we compute their \textit{relative mutual information} (RMI) with the class~\cite{cover1991entropy}.
RMI is a measure of how good the feature is for discriminating between samples that belong to different classes.
For a feature whose distribution is $x$, RMI is computed as the percentage difference between the marginal entropy of the feature distribution $H(x)$ and its conditional entropy given the class label $H(x|y)$, with respect to the initial entropy $H(x)$, that is:
\[ RMI(x, y) = \frac{H(x) - H(x|y)}{H(x)}. \]
For the quantization, we use 256 linearly distributed bins among the minimum and the maximum of the distribution.

\begin{figure}[htpb]
    \centering
    \includegraphics[width=0.45\textwidth]{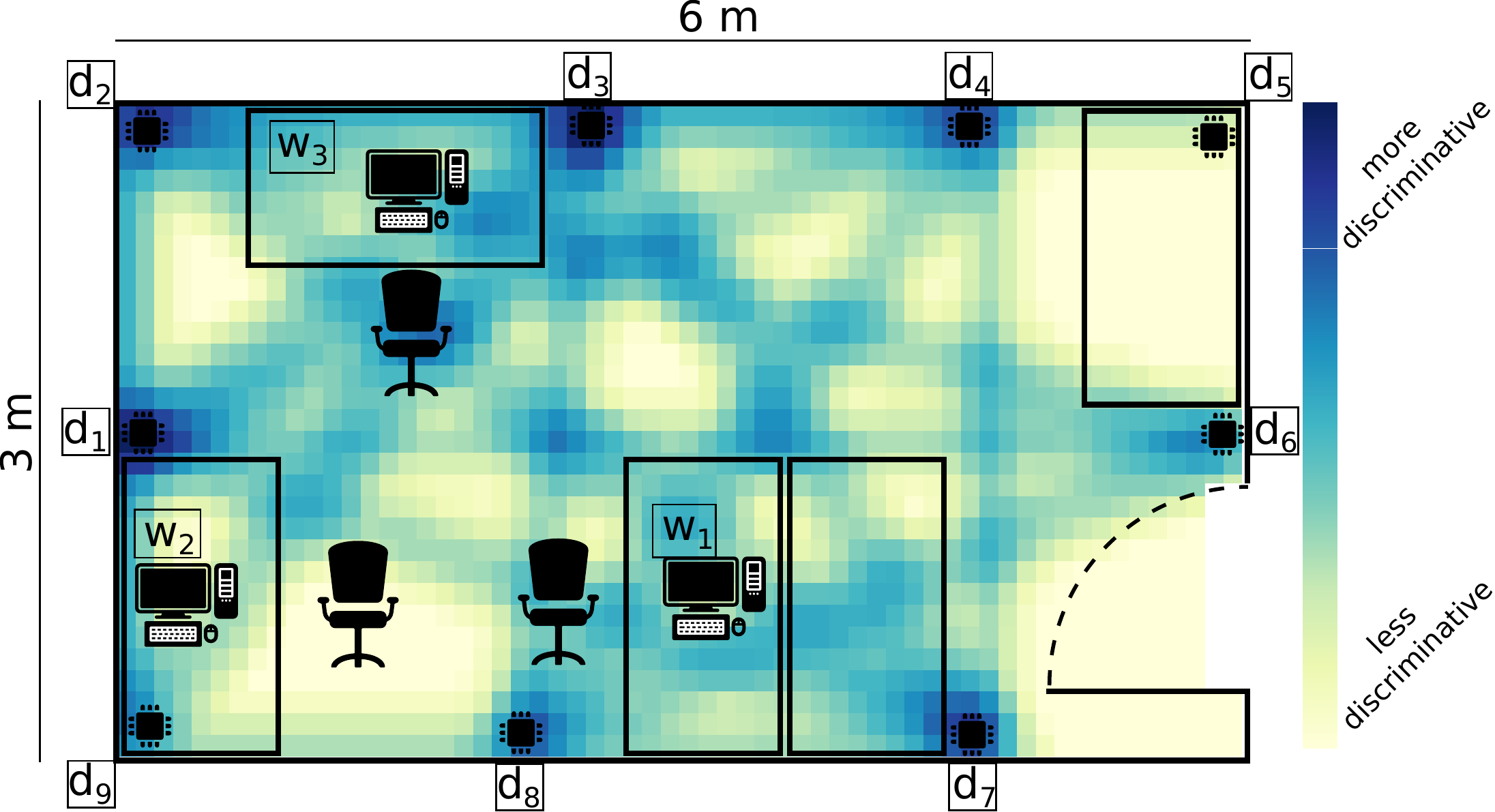}
    \caption{Importance of the streams in terms of RMI visualized as an heatmap on the planimetry of the office used for the experiment.}\label{fig:setup_heatmap}

\end{figure}

With the RMI values, we plot an heatmap of the importances of the single streams between devices, reported in Figure~\ref{fig:setup_heatmap}.
In the figure, darker areas correspond to stronger importance for the features of the streams that pass through that area, in terms of RMI.
As shown, certain devices (e.g., $d_5$) do not significantly contribute to the classification, as the their information is either not discriminative, or is strongly correlated with other devices (and is therefore not colored in Figure~\ref{fig:setup_heatmap}).

In Table~\ref{tab:features_importances}, we report the 15 features that scored higher in terms of  RMI in our data in.
The feature names indicate the stream, and the type of feature for that stream, either entropy (\textit{ent}), variance (\textit{var}), or autocorrelation (\textit{ac}).

\begin{table}[htpb]
\begin{center}
\begin{tabular}{c|c|c}
   \textbf{Rank} & \textbf{feature} & \textbf{RMI} \\
\hline \textbf{1} & d9-d2-ent & 0.2977 \\
\hline \textbf{2} & d7-d8-ac & 0.2863 \\
\hline \textbf{3} & d7-d1-ent & 0.2858 \\
\hline \textbf{4} & d1-d3-ac & 0.2809 \\
\hline \textbf{5} & d4-d2-ac & 0.2807 \\
\hline \textbf{6} & d3-d5-ac & 0.2778 \\
\hline \textbf{7} & d6-d8-ac & 0.2776 \\
\hline \textbf{8} & d3-d9-ac & 0.2770 \\
\hline \textbf{9} & d6-d2-ac & 0.2765 \\
\hline \textbf{10} & d4-d1-ac & 0.2761 \\
\hline \textbf{11} & d2-d3-ac & 0.2757 \\
\hline \textbf{12} & d1-d9-ac & 0.2752 \\
\hline \textbf{13} & d1-d6-var & 0.2711 \\
\hline \textbf{14} & d8-d9-ac & 0.2707 \\
\hline \textbf{15} & d8-d4-ac & 0.2698 \\
\end{tabular} 
\caption{RMI for the top 15 ranking features.}
\label{tab:features_importances}
\end{center}
\end{table}

\subsection{Comparison}\label{sec:comparison}

\begin{figure}[htpb]
    \centering
    \includegraphics[width=0.45\textwidth]{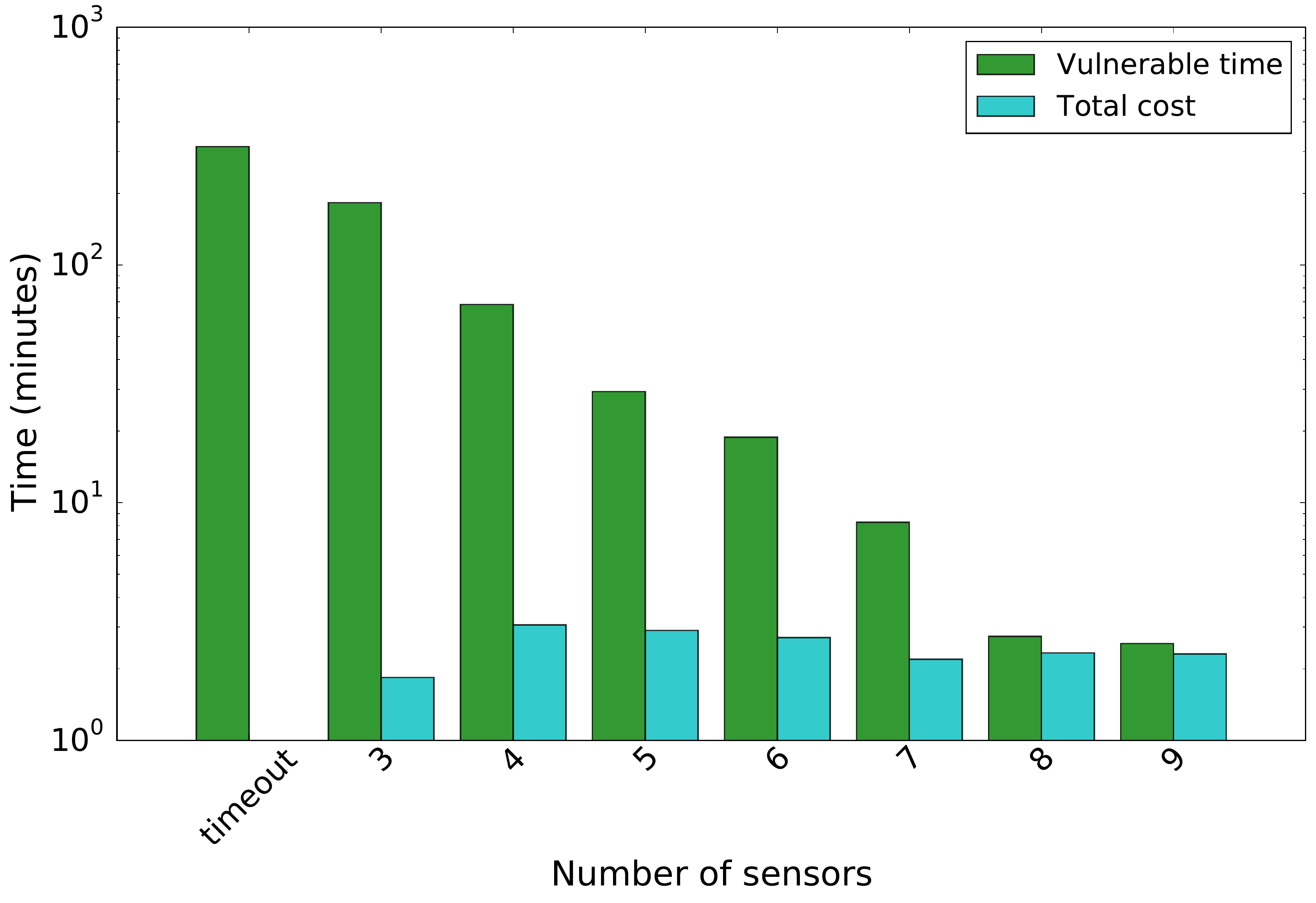}
    \caption{Comparison between the vulnerable time for the workstations and the total cost for the users, for an approach with time-out, and increasing number of sensors.}\label{fig:comparison}
\end{figure}

In order to evaluate the usability and security aspects side by side, we introduce a broader indicator for the security, that is the amount of time workstations spend in a \textit{vulnerable} state (i.e., unattended and authenticated).
In fact, in our approach, we are reducing the time that workstations spend in a vulnerable state, while increasing the cost for the users.
Figure~\ref{fig:comparison} shows such a trade-off, comparing the time-out approach (with $T=300$ seconds time-out) with the outcome of the system with increasing number of sensors.
As shown in the figure, with a time-out approach there is no cost for the users, meanwhile, increasing the number of sensors, the cost increases as well (system is querying users and possibly committing errors in the deauthentication).
However, although the cost for the users initially increases, it stabilizes soon for an increasing number of sensors (after four sensors in our experiment).
Furthermore, the increment in cost is compensated by an exponential decrement in the vulnerable time.
This means that increasing the number of sensors does not add significant burden for the users, while it brings valuable improvements in terms of  security.

\end{document}